\documentclass[fleqn,10pt,showpacs,twocolumn]{revtex4}

\usepackage{amsmath}
\usepackage{amssymb}
\usepackage{graphicx,psfrag,epsfig}

\newcommand{\be}{\begin{equation}}
\newcommand{\ee}{\end{equation}}
\newcommand{\bea}{\begin{eqnarray}}
\newcommand{\eea}{\end{eqnarray}}
\newcommand{\bean}{\begin{eqnarray*}} 
\newcommand{\eean}{\end{eqnarray*}}

\newcommand{\bm}[1]{\mbox{\boldmath $#1$}}

\newcommand{\ket}[1]{\vert #1 \rangle}

\def\!{\hat}
\def\slash{\rlap{/}}

\def\slash#1{\setbox0=\hbox{$#1$}               
        \dimen0=\wd0                            
        \setbox1=\hbox{/} \dimen1=\wd1          
        \ifdim\dimen0>\dimen1                   
        \rlap{\hbox to \dimen0{\hfil/\hfil}}    
        #1                                      
        \else                                   
        \rlap{\hbox to \dimen1{\hfil$#1$\hfil}} 
        /                                       
        \fi}                                    %

\begin{document}

\title{Emergent symmetries of the Standard Model}

\author{P.J.~Mulders}
\email{p.j.g.mulders@vu.nl}
\affiliation{
Nikhef Theory Group and Department of Physics and Astronomy, 
VU Amsterdam,\\
De Boelelaan 1081, NL-1081 HV Amsterdam, the Netherlands}

\begin{abstract}
We show how, using multipartite entanglement, the symmetries among bosons and fermions of the Standard Model of particle physics emerge. Fermions belong to tripartite maximally entangled classes starting with basic chiral right and left states. Quarks and leptons belong to different classes, with only leptons appearing as asymptotic states in three space dimensions. The Higgs boson is the scalar mode with a nonvanishing vacuum expectation value, other bosons are as usual linked to symmetry generators.  
\end{abstract}
\date{\today}
\pacs{11.30.Cp, 12.10.Dm, 12.15.-y, 12.38.Aw}

\maketitle

\section{Introduction}

The Standard Model (SM) of particle physics encompasses our knowledge and understanding of the elementary particles and forces and successfully describes the matter in the 3D world, where 3D refers to the number of space dimensions. The theory is built around the SU(3)$_C \otimes$[SU(2)$\times$U(1)]$_{EW}$ gauge symmetry with unbroken color (C) symmetry and electroweak (EW) symmetry spontaneously broken down to the electromagnetic U(1)$_{EM}$ symmetry. It exhibits a left-right (L-R) chiral symmetry as well as the P(1,3) Poincar\'e symmetry of space-time. The spectrum in 3D has elementary fermionic and bosonic modes. The Higgs field is the scalar bosonic mode in the theory with a nonzero vacuum expectation value (vev). Other bosonic modes are associated with the symmetry generators, massive ones with generators of spontaneously broken symmetries. Fermionic modes are leptons with EW interactions and quarks with both EW and color interactions. In 3D most signs of color are hidden (confinement), neither is there an obvious fermion-boson supersymmetry (SUSY).

In this letter, we restrict ourselves to the emergence of quantum states and the relevant symmetries in the SM. For this we borrow concepts from Quantum Information Theory (QIT). One of these concepts are multipartite states. Entanglement is a basic property of multipartite quantum states and one distinguishes classes of entangled states. Take as the basic starting point a qubit basis of right (R) and left (L) states, $\ket{R}$ and $\ket{L}$ in the Hilbert space ${\cal H}$. For bipartite states in ${\cal H}\otimes{\cal H}$ there is just a single class of entangled states, the Bell-states with as a maximally entangled representant $\ket{\rm Bell} = \tfrac{1}{\sqrt{2}}(\ket{RR} + \ket{LL})$ or equivalently the states $\tfrac{1}{\sqrt{2}}(\ket{RL} \pm \ket{LR})$. These representing states are equivalent under local unitary (LU) transformations $V_A\otimes V_B$, where local refers to any of the subspaces. More precisely the equivalence for entanglement classes is under stochastic local operations and classical communication (SLOCC). For details we refer to Ref.~\cite{Dur:2000zz}, in which a thorough study is presented of the entangled tripartite states in ${\cal H}^{\otimes 3} = {\cal H}\otimes{\cal H}\otimes{\cal H}$. Tripartites belong to two classes represented by the GHZ and W-states~\cite{GHZ:1999,Bouwmeester:1998iz,Dur:2000zz,Walter:2015},
\bea
&&
\ket{{\rm GHZ}} = (\ket{RRR} + \ket{LLL})/\sqrt{2},
\label{GHZ}
\\&&
\ket{{\rm W}} = (\ket{LRR} + \ket{RLR} + \ket{RRL})/\sqrt{3}.
\label{W}
\eea
Just as Bell states, these states are maximally entangled (MaxEnt) states. Entanglement is measured by looking at the non-purity of the reduced density matrix $\rho$. It is measured by $0\le 1- {\rm Tr}(\rho^2)\le 1$ (0 for pure systems) or by quantities like the von Neumann entropy $S = - {\rm Tr}(\rho\ln\rho)$. As will be shown in the remainder of this letter, these concepts can in combination with basis sets of two and three degrees of freedom be used to understand how all symmetries of the SM emerge. Preliminary ideas on this have been outlined in Refs.~\cite{Mulders:2016qve,Mulders:2016ofb,Mulders:2018mmg}.

For the study of emergence of symmetries, we note that symmetry eigenstates are in general entangled. E.g.\ allowing all phases, the bipartite qubit states belonging to ${\cal H}\otimes {\cal H}$ can be organized in singlet and triplet representations of SU(2)$_T$, with generators $\bm T = \bm t\otimes\bm 1 + \bm 1\otimes\bm t$. All states are built from product states $\ket{a}\otimes\ket{b} = \ket{ab}$ and have a density matrix with maximal rank two (using a Schmidt decomposition). The mixed symmetry states $\ket{RL}\pm\ket{LR}$ are MaxEnt. The aligned symmetry states $\ket{RR}$ and $\ket{LL}$ are not entangled and actually form a doublet under diagonal transformations $V\otimes V$ generated by $\bm t\otimes\bm t$, which we will refer to as aligned fundamental representations. The aligned symmetry states can be combined into the MaxEnt state $\ket{RR} + \ket{LL}$.
Our main conjecture is that the physically relevant degrees of freedom correspond to the classes of MaxEnt tripartite states. For this we look at the equivalence classes of entangled states, their symmetries and the reduced density matrices, following the QIT methods of Ref.~\cite{Dur:2000zz}. We note that the selection of particular suitable bases of entangled multipartite states can play a role in classical versus quantum behavior~\cite{Hooft:2014kka} and might even be governed by a MaxEnt principle~\cite{Cervera-Lierta:2017tdt} as a rule. 

\section{\label{section_basic} Basic states}

\begin{figure}[b]
\epsfig{file=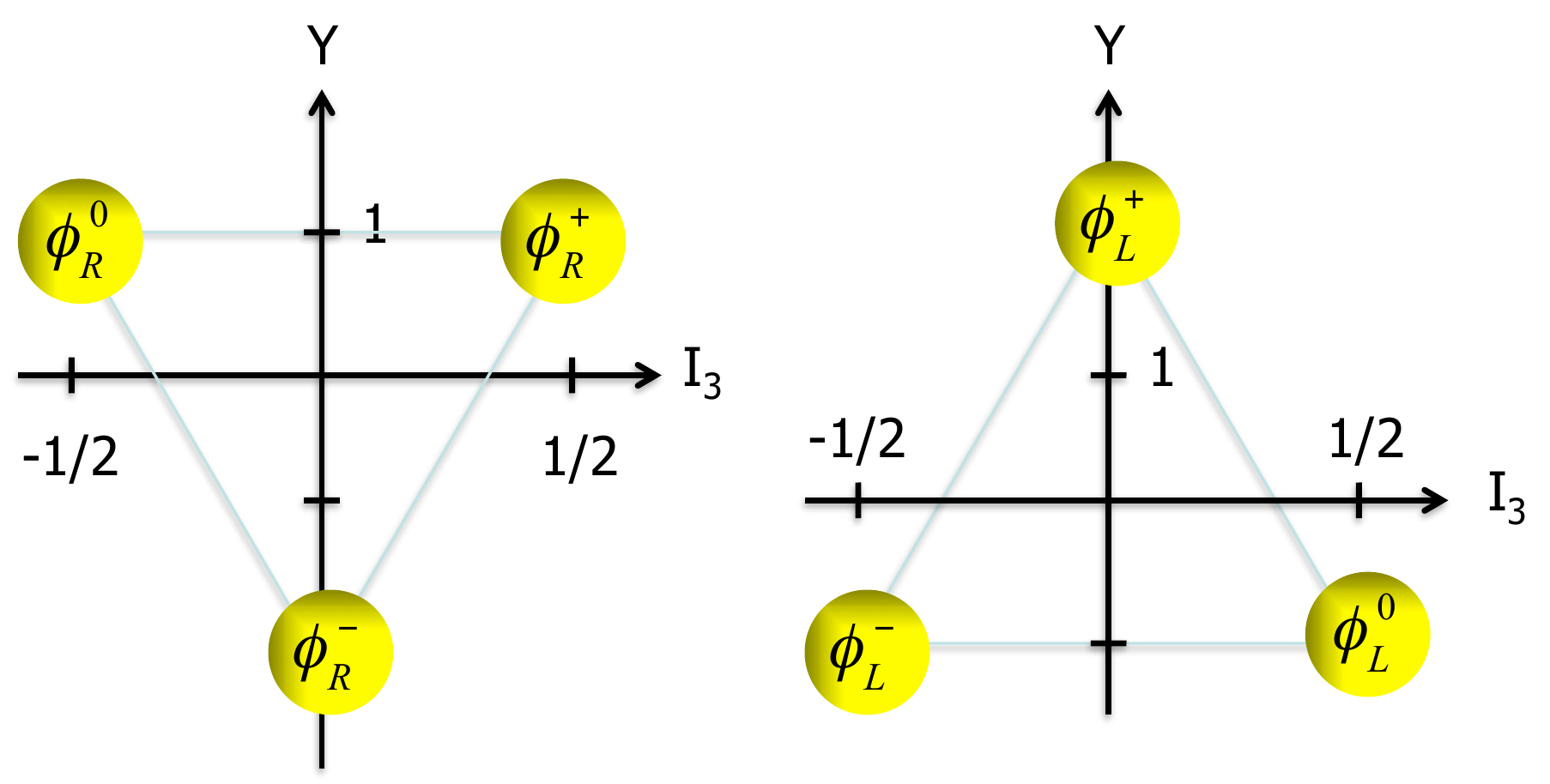,width=0.35\textwidth}
\epsfig{file=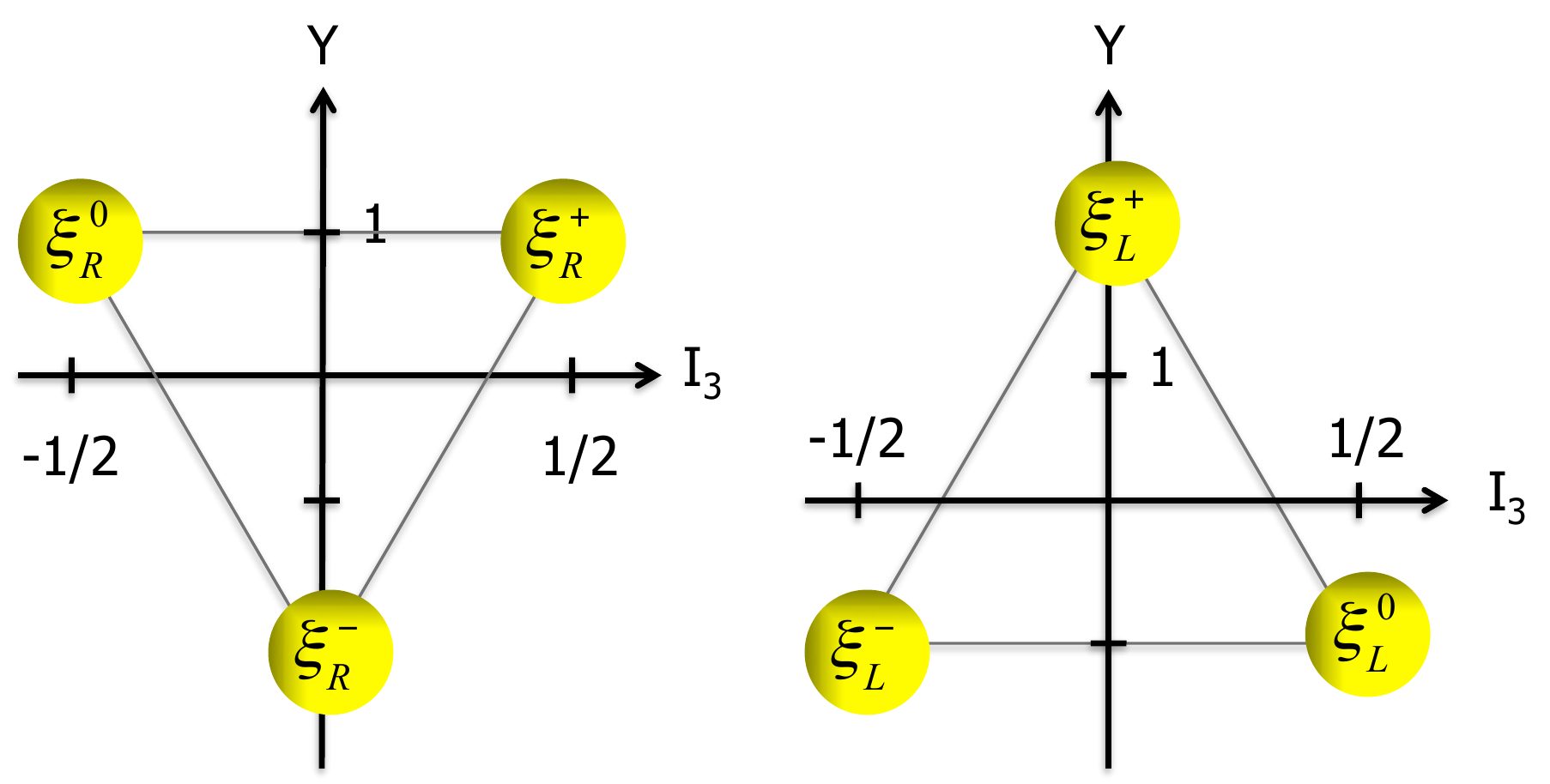,width=0.35\textwidth}
\caption{\label{basic-ch}
Basic triplet/antitriplet of right-handed ($\phi_R^i$) and left-handed bosons ($\phi_L^i$) and the corresponding fermion multiplets. 
Choosing electroweak axes ($Y$, $I_3$) and the charge assignment $Q = I_3 + Y/2$ they are identified as charged and neutral states,
e.g.\ for bosons $\phi_R^+$, $\phi_R^0$, $\phi_R^-$.
} 
\end{figure}

In order to reproduce the SM symmetries for tripartite states, it is sufficient to start with a much simpler basis in the subspace. One needs right (R) and left (L) states and in addition three real degrees of freedom labeled $i = 1, 2, 3$, which in the usual way can be reorganized into a complex charged pair (charges $\pm$) and one neutral degree of freedom. The relevant symmetries in the Hilbert space are local unitary SU(3) rotations, under which the right states form a triplet ($\underline 3$) representation and the left states form an anti-triplet ($\underline 3^\ast$) representation, such that the basis is CP-symmetric under the combination of charge conjugation (C) and R-L interchange (parity P). Discrete symmetry groups of the qubits R/L and the triplet basis are Z(2) and Z(3), subgroups of A(4), algebraically represented by generators $S$ for Z(2) and $T$ for Z(3) satisfying $S^2 = T^3 = (ST)^3 = 1$. To get the right symmetries of the SM one finally needs bosons and fermions as basic states in a supersymmetric fashion, shown in Fig.~\ref{basic-ch}.

The degrees of freedom are described with hermitean fields that create/annihilate the states, $\varphi = ( a + a^\dagger)/\sqrt{2\omega}$ and $\xi = (b + b^\dagger)/\sqrt{2}$ with $[a,a^\dagger] = \{b,b^\dagger\} = 1$, where the excitations $\varphi$ for the boson fields are around a bosonic vev. It is natural to look at entanglement classes for fields as these are sums over allowed modes. The anti-hermitean supercharge combination $Q = \sqrt{\omega}(b^\dagger a - b a^\dagger)$ transforms fermions into bosons and vice versa, satisfying 
\bea
&&
\{Q,Q^\dagger\} = \omega(\{a^\dagger,a\} + [b^\dagger,b]),
\label{QQ}
\eea
which contains on the RHS the Hamiltonian, number operator multiplying energy scale $\omega$. In the presence of internal degrees of freedom, there are several Hamiltonians (diagonal entries on the RHS) but it also incorporates all (unitary) transformations in internal space. The algebra for the fields is
\bea
&&
[Q,\phi] = \xi,
\\
&& F = \{Q,\xi\} = \{Q,[Q,\phi]\} = iD\phi,
\label{basicDb}\\
&& [Q,F] = [Q,\{Q,\xi\}] = iD\xi,
\label{basicD}
\eea
where the covariant derivative $iD = i\partial + gA$ splits into a part $i\partial$ governing space-time dependence linked to the Hamiltonians and gauge fields $gA$, emerging from the pseudoscalar degrees of freedom, that ensure local symmetry, which in our multipartite setting includes space-time {\em and} internal transformations.

\section{Space-time dependence and local gauge invariance}

With the basic starting point being a single degree of freedom, there is only a single Hamiltonian on the RHS of Eq.~\ref{QQ}, which describes time-dependence as an implementation of the U(1) symmetry via Eq.~\ref{basicDb}, with the RHS just being $iD\phi = \dot\phi$. With decoupled right- and left degrees of freedom one has two (commuting) Hamiltonians, $P^\pm = H \pm P$, allowing (light-cone) times $x^\pm$ as implementation of the U(1)$_R\otimes$U(1)$_L$ symmetry. Coupling R/L and including the boost operator, $[K,P^\pm] = \pm i\,P^\pm$, one obtains the Poincar\'e P(1,1) algebra with Casimir operator $M^2 = P^+P^-$. The space-time dependence emerges through the full covariant derivative,
\bea
&&
\phi(x) = {\cal P}\exp\left(-i\int_0^x ds{\cdot}D(s)\right)\phi,
\eea
where the field sums over momentum modes. Provided that all symmetries are present in Minkowski space {\em and} field space, this establishes the Poincar\'e invariance of fields. For instance, for the P(1,1) symmetry, a pure gauge $gA_\sigma^{\rm pure} = i\,\partial_\sigma\eta$ corresponds to a boost with rapidity $\eta$, bringing $\chi \rightarrow \chi\,e^{\eta}$ and $(\tau,0) \rightarrow (\tau\cosh\eta,\tau\sinh\eta)$. One does need this matching symmetry in field space in order to respect the Coleman-Mandula theorem~\cite{Coleman:1967ad}. This is basically what is the situation for the quite general Wess-Zumino (WZ) lagrangian~\cite{Wess:1973kz}, which for our purposes would be restricted to 1+1 dimension. In this lagrangian the boost symmetry is present through field constraints $2\phi_R\phi_L = \phi_S^2-\phi_P^2 = 1$, the latter containing the scalar and pseudoscalar fields  $\phi_{R/L} = (\phi_S\pm\phi_P)/\sqrt{2}$. The pseudoscalar field also serves as the gauge field in 1D. The WZ lagrangian has a mass $M$ for fermions and bosons that not only couples left and right but also serves as the dimensionful fermion-boson coupling $g = M/2$. 
Besides pure gauge transformations, physical effects of gauge symmetries are governed by the Wilson loop $W[C] = \exp(-ig\oint ds{\cdot} A)$ or equivalently locally nonvanishing field strength $gF_{\tau\sigma} = \delta W[C]/\delta \sigma^{\tau\sigma}$. 
With the WZ action in 1D one has a full supersymmetric field theory with two dimensionless bosonic fields, recoupled to a single (complex) field
\bea
\phi\sqrt{2} &=& e^{i\pi/4}\phi_R + e^{-i\pi/4}\phi_L = \phi_S + i\phi_P,
\label{Lfield-complex}
\eea
with nonzero vev $\langle\phi_S\rangle = 1$, and two Majorana fields combined into a two-component (chiral) fermionic field,
\bea
\psi &=& \frac{1}{\sqrt{2}}\left[\begin{array}{c}\xi_R \\ -i\xi_L\end{array}\right].
\eea
Returning to the space-time symmetries with a mass term coupling R and L the Hamiltonian becomes
\bea
&&
H = \alpha\,P + \beta\,M = \left[\begin{array}{cc} P & M \\ M & -P\end{array}\right],
\label{Lhamiltonian11}
\eea
with $\beta = \gamma_0 = \rho^1$ off-diagonal and $\alpha = \gamma_0\gamma^3 = \rho^3 = \gamma_5$ being diagonal (+1, $-1$). The gamma matrices $(\gamma^0,\gamma^3) = \beta(1,\alpha) = (\rho^1,-i\rho^2)$ satisfy $\{\gamma^\rho,\gamma^\sigma\} = 2\,g^{\rho\sigma}$, explicitly
\bea
&&
\gamma^\rho = \left[\begin{array}{cc} 0 & n_-^\rho \\ n_+^\rho & 0 \end{array}\right],
\ \gamma_5 = \left[\begin{array}{cc} 1 & 0 \\ 0 & -1 \end{array}\right],
\eea 
which is all well-known when implementing covariance and in particular boost invariance for fermions. 
   
The step from 1D to 3D involves extending the Hamiltonians from $(H,P) \rightarrow (H,\bm P)$ using $SO(3)$ invariance, the P(1,3) algebra being the closure of the commutator algebra [P(1,1),SO(3)]. This rotates $(n^0,n^3) \rightarrow (n^0,\bm n)$ and for fermions take $(1,\pm 1) \rightarrow (1,\pm\bm\sigma)$ to get the 1+3 Clifford algebra (in chiral representation) for good light-front fermion fields~\cite{Dirac:1949cp,Kogut:1969xa}. The 1D pseudoscalar fields are combined into 3D vector fields. For all fields one has the standard requirements of Poincar\'e invariance, while an action that allows curvature in the 1D to 3D extension is possible~\cite{Jacobson:2003vx}. 
The 1D to 3D extension is the tripartite entanglement of three real degrees of freedom, analogous to the 3D harmonic oscillator (HO) viewed as entangled tripartite quantum states in the product space of three 1D HOs with the SU(3) invariance still being visible in the spectrum~\cite{Georgi:1999wka}. It is important to realize that the Z(3) symmetry among the three space directions given the time direction and the Z(2) symmetry associated with R-L, implies an oriented embedding of E(1,1) in E(1,3) space-time, governed by the A(4) symmetry that was already mentioned in the previous section.

\section{\label{tri} Tripartite entanglement}

\begin{figure}[t]
\epsfig{file=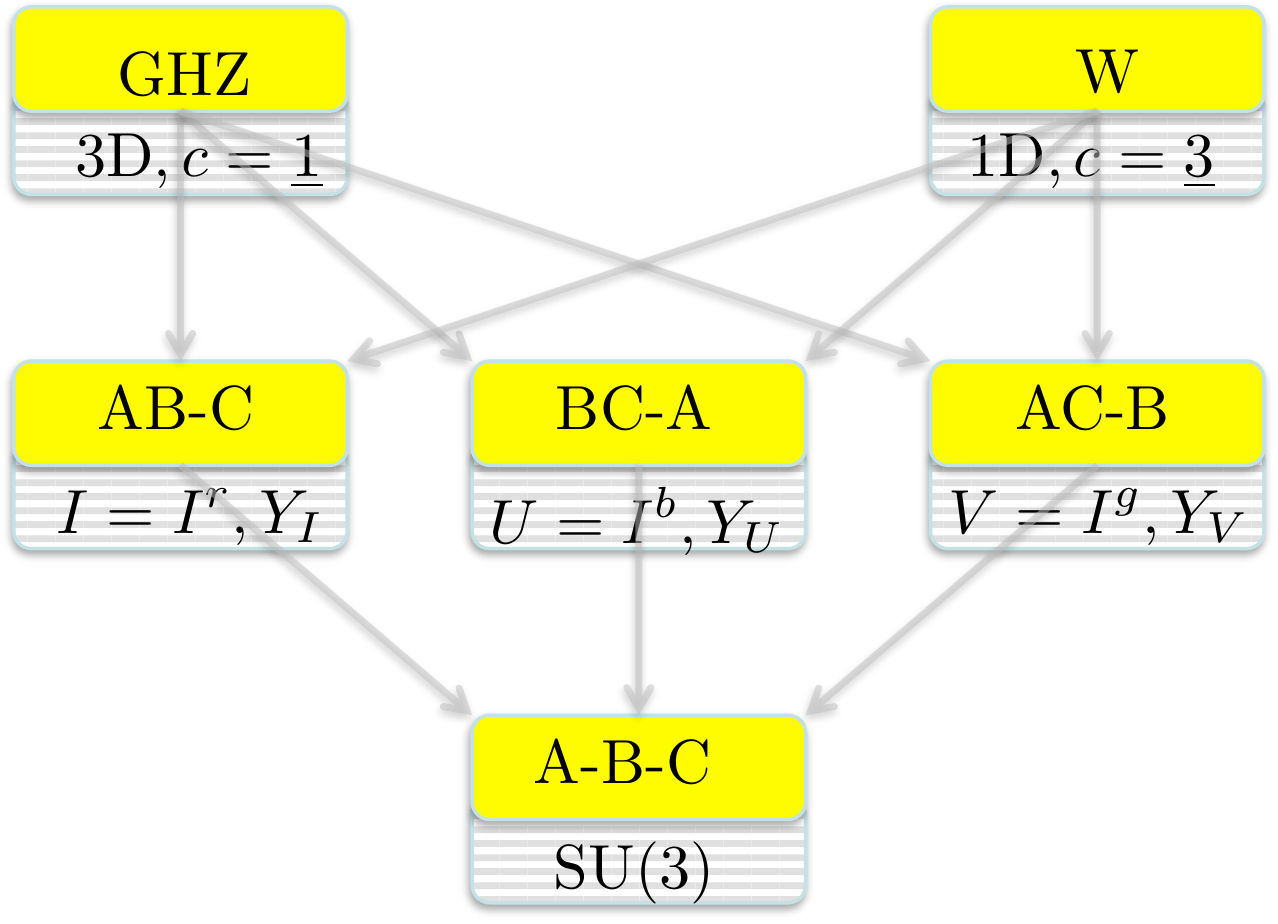,width=0.35\textwidth}
\caption{\label{tripartite}
Different classes of local MaxEnt tripartite pure states. The arrows indicate the action of, in general non-invertible, local operators between classes of entangled states and the underlying product states (Ref.~\cite{Dur:2000zz}, Fig.~1). Added to the classes are symmetries discussed in Section~\ref{tri} (3D or 1D, color, I-, U- or V-spin).
} 
\end{figure}

Looking for MaxEnt SLOCC-equivalent classes for tripartite states, only the R-L qubit structure matters. Because of LU the SU(3) symmetry is not relevant for entanglement. As pointed out in Ref.~\cite{Dur:2000zz}, it is useful to consider the tripartite-entangled GHZ and W classes together with the three classes of product states involving bipartite-entangled states. We label these three classes I, U and V, because with the assignment of $I^3-Y$ the AB-C class has entangled I-spin symmetry eigenstates or aligned combinations of them. The two other classes contain U-spin or V-spin symmetry states. With this assignment shown in Fig.~\ref{tripartite} for MaxEnt states, we can study the tripartite modes built from the bases in Fig.~\ref{basic-ch}. 

\subsection{\label{4a} Leptons}

\begin{figure}[t]
\epsfig{file=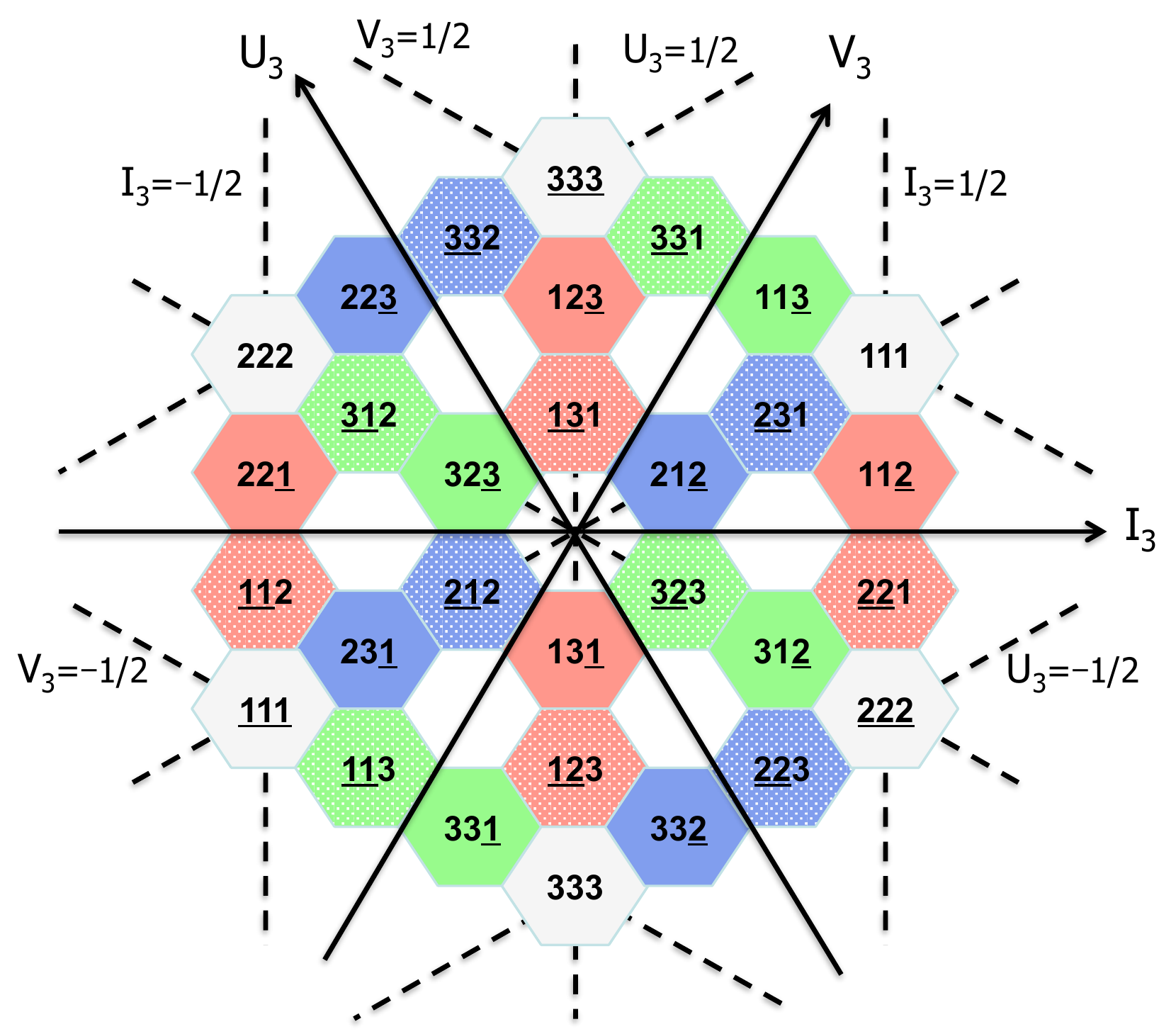,width=0.45\textwidth}
\caption{\label{Lroots-euclidean}
Roots for $SU(3)$ that are allowed for GHZ- and W-type tripartite states made up of right (1,2,3) or left (\underline 1, \underline 2, \underline 3) states. Using standard $I^3-Y$ assignment, states with integer or half-integer values of $I$-, $U$- or $V$-spin are shown. The actual values of the roots for tripartite states are represented as averages, $I^3 = (I^3_A + I^3_B + I^3_C)/3$, etc.  
} 
\end{figure}

The easiest states to analyse are the GHZ fermions, which involve qubits RRR or LLL in the fundamental aligned SU(3) representation, the same as the original basis. The reduced density matrix of $\rho = \vert {\rm GHZ}\rangle\langle {\rm GHZ}\vert$, 
\be 
\rho_{AB} = {\rm Tr}_C(\rho) = \tfrac{1}{2}\left(\vert RR\rangle\langle RR\vert + \vert LL\rangle\langle LL\vert\right),
\ee 
has rank 2 and is a mixture of pure states involving aligned I-spin symmetry states in AB-C class. The other reduced matrices involve aligned U-spin and V-spin states. The easiest way to look at the symmetry structure of the fermions is to look at the full set of SU(3) roots, using the well-known I-, U- or V-spin axes. This is done for tripartite states starting with the basis states in Fig.~\ref{basic-ch}. Actually starting with a symmetric basis, the results are shown in Fig.~\ref{Lroots-euclidean} as I-, U- or V-spin states. The  GHZ fermions correspond to the corners in the diagram in Fig.~\ref{Lroots-euclidean} having allowed roots for I-spin, U-spin as well as V-spin. At this point, it is useful to realize that the discrete A(4) $\subset$ SU(3) symmetry implies that the transformations $\bm t_I\otimes \bm t_U\otimes \bm t_V$ are LU equivalent to diagonal transformations $\bm t_I\otimes \bm t_I\otimes \bm t_I$, where $\bm t_I = (\bm I,Y_I)$ are the generators of the relevant SU(2)$_I\times$U(1)$_Y$ in the AB-C bipartite entangled class, etc. Since A(4) is a subgroup for both SO(3) and SU(3), one can label the aligned tripartite states as SO(3) representations and include this group into a P(1,3) symmetry group, forming a direct product P(1,3)$\otimes$[SU(2)$_I\times$U(1)$_Y]$. In this way the aligned GHZ fermions are identified as leptons, shown in Fig.~\ref{Lexcitations-electroweak}. These leptons have integer charges and live in 3D. The leptons belong to singlet representations of the A(4) embedding group, which has three singlet and three triplet representations. This allows the identification of three families of leptons where the A(4) symmetry plays an important role in the family mixing~\cite{Cabibbo:1977nk,Wolfenstein:1978uw,Ma:2001dn,Ma:2004zv,Altarelli:2005yx,Harrison:2002er}. 

\subsection{\label{4b} Quarks and confinement}

\begin{figure}[t]
\epsfig{file=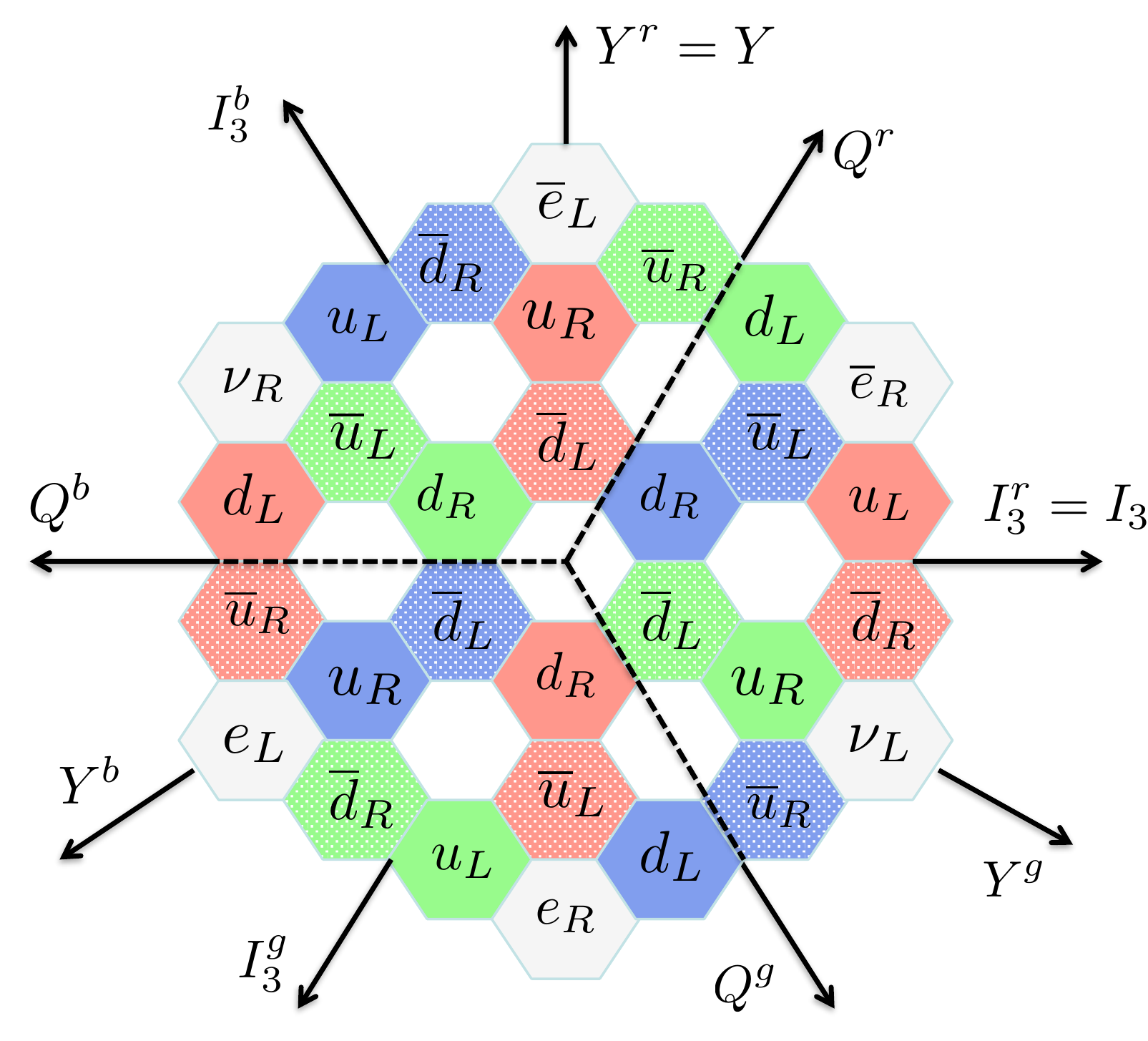,width=0.39\textwidth}
\caption{\label{Lexcitations-electroweak}
Excitations and assignment of electroweak quantum numbers for both leptons and colored quarks as GHZ- and W-classes of tripartite entangled states. The charges of tripartite states are $Q = I^3 + Y/2$ for leptons and $Q = I_3^{r}+Y^r/2 = I_3^{g}+Y^g/2 += I_3^{b}+Y^b/2$ for quarks.
} 
\end{figure}

The second class of entangled tripartite states are W-states with R-L entanglement given in Eq.~\ref{W}. The reduced density matrix of $\rho = \vert {\rm W}\rangle\langle {\rm W}\vert$ is 
\be 
\rho_{AB} = {\rm Tr}_C(\rho) = \tfrac{2}{3}\left(\vert {\rm Bell}\rangle\langle {\rm Bell}\vert + \vert RR\rangle\langle RR\vert\right),
\ee 
which has rank 2 but remains entangled in the AB-C class of I-spin states. Similarly the other reduced density matrices remain entangled in bipartite U-spin or V-spin classes. Identifying $\bm t_I\otimes \bm t_U\otimes \bm t_V$ as $\bm t^r\otimes \bm t^b\otimes \bm t^g$ (with colors r, g, b) shows that there is a symmetry structure SU(3)$_C\times$[SU(2)$_I\times$U(1)$_Y]$ where the electroweak W-states also belong to triplet (or anti-triplet) representations of SU(3)$_C$, leading to the natural identification with colored quarks and antiquarks. With respect to the A(4) symmetry group, the quarks also form triplets, leading also for quark (W) classes to a three-family structure, one of them massive. The EW structure of the quarks is illustrated in the root diagram in Fig.~\ref{Lroots-euclidean}, in which the W-states appear as I-, U- {\em or} V-spin states with an exclusive {\em or}. This does have consequences for the electroweak, including charge, assignments of the quarks as compared to those of the leptons, shown in Fig.~\ref{Lexcitations-electroweak}.  But, remarkably, for each color it leads to a left quark doublet and a right antiquark doublet as well as right and left singlets. The way in which the EW structure emerges resembles the rishon model~\cite{Harari:1979gi,Shupe:1979fv,Harari:1981uh}, but rather than having two fractionally charged preons ($V$ and $T$) in 3D, the basic modes are the charged or neutral 1D excitations, evading the necessity of compositeness.
Taking the W-states to belong to representations of an unbroken SU(3)$_C$ symmetry replaces the implementation of the SO(3) symmetry in the tripartite space. Thus one has current quarks or partonic quarks living in 1D with just a P(1,1) space-time symmetry. 

It is possible to construct quark-entangled hadron states in 3D (see e.g.~\cite{Kharzeev:2017qzs}). Since Z(3)$\subset$SO(3)$\subset$SU(3) it is sufficient to construct global SU(3)$_C$ singlets, fully antisymmetric in tripartite space. Confinement has then been rephrased as a choice of the class of tripartite states with the appropriate symmetries. The quarks are entangled into symmetric chiral and spatial 'valence' modes, composite in space, where baryons are the color-antisymmetrized combinations of these 'valence' modes (see also Ref.~\cite{Zenczykowski:2008xt}). 

\subsection{\label{4c} Bosons}

\begin{figure}[t]
\epsfig{file=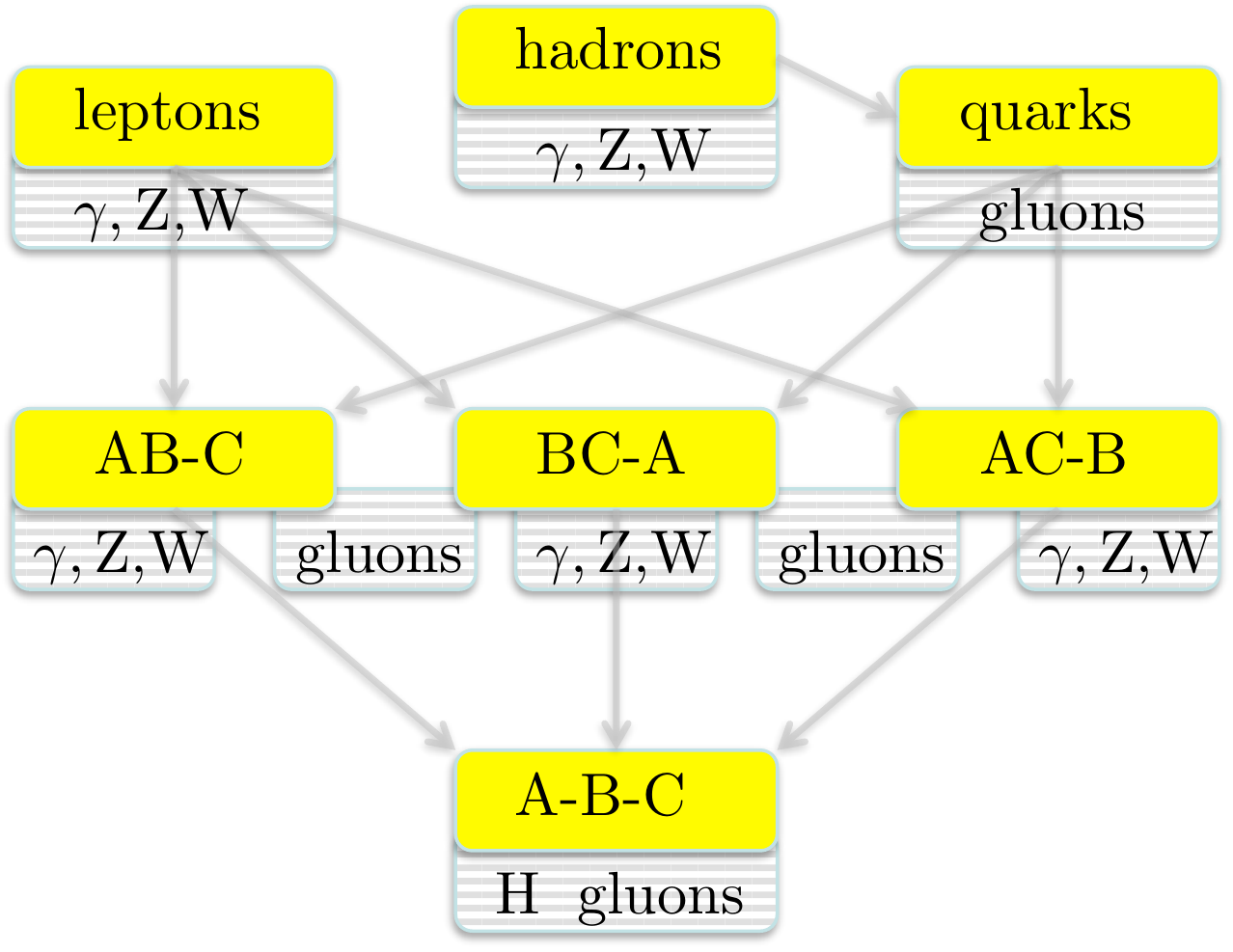,width=0.35\textwidth}
\caption{\label{tripartite-physics}
Identification of MaxEnt classes with physical bosonic degrees of freedom linked to the symmetries in the classes (cf.\ Fig.~\ref{tripartite}). Hadrons are added as composite states involving entanglement in space, discussed in Section~\ref{4b}.
} 
\end{figure}

The relevant symmetry for the fields without a mass term is U(1)$_R\times$SU(3)$_R\times $U(1)$_L\times$SU(3)$_L$, while with a mass term the symmetry becomes P(1,1)$\otimes $SU(3), where the $P(1,1)$ includes a discrete Z(2) parity symmetry and $SU(3)$ includes the full range Z(2)$\times$Z(3)$\subset$A(4)$\subset$SO(3)$\subset$SU(3). The vacuum is R-L symmetric and invariant under P(1,1)$\times$SO(3)$\times$U(1)$_{EM}$.

Even if for basis states (Fig.~\ref{basic-ch}) bosons mirror the fermions, their role in the entangled multipartite spaces is different. Writing $\phi\sqrt{2} = \chi\,e^{i\theta}$, the coupling of bosons to fermions is different for the scalar field $\chi$ and the pseudoscalar phase(s) $\theta$, depending on the structure of the action, the constraints implied by the action and the vev of the bosonic fields. In tripartite space with a triplet of complex fields $\vec\phi = (\phi^1,\phi^2,\phi^3)$ and an SO(3) symmetric vev, $\vert\vec\phi\vert = 1$ or $\langle\vec\phi\,\rangle = \hat\phi$, the gauge freedom can be used to eliminate the angles giving $\vec\chi = (1+\varphi_H)\hat\phi$ where the fluctuation $\varphi_H$ corresponds to the scalar Higgs mode. The fields $\theta$ ends up in the gauge fields by writing
\be 
\tfrac{1}{2}\,\chi^{T} D_\sigma \chi = \phi^{\dagger}\partial_\sigma \phi,
\ee
replacing in 1D the pseudoscalars $\phi_P$ by eight (instantaneous) gauge fields for the $G = SU(3)$ symmetry, carrying color charge $c = \underline 8$, and leaving in 1D just an innocent scalar mode resembling XQCD as in Ref.~\cite{Kaplan:2013dca}. In 3D one has four vector gauge fields for the $G^\prime = SU(2)\times U(1)$ symmetry with electroweak charges $(I_W,Y_W) = (1,0)$ and $(0,0)$. They take care of the local coupling to fermions, local both in tripartite space as well in space-time. These gauge fields appear as discussed at the end of Section~\ref{section_basic} in the covariant derivatives,  
\bea
\mbox{1D}:&&  
i D_\sigma\psi^i = i\partial_\sigma\psi^i + \sum_{a\in \underline G} gA_\sigma^a (T_a)^i_j\psi^j,
\label{covfderL1}
\\
\mbox{3D}:&&
iD_\mu\psi^i = i\partial_\mu\psi^i 
+ \sum_{a\in \underline G^\prime} gA_\mu^a (T_a)^i_j\psi^j.
\label{covfderL3}
\eea
The bosonic quantum states have a bipartite RL structure and, belonging to definite representations of the symmetries, form bipartite entangled Bell states fitting in the classes AB-C, etc.\ of Figs~\ref{tripartite} or \ref{tripartite-physics}. The nonet of 3D gauge bosons is LU equivalent to aligned RR and LL symmetry states, $\underline 3 \times \underline 3 = \underline 6 + \underline 3^\ast$ and $\underline 3^\ast \times \underline 3^\ast = \underline 6^\ast + \underline 3$. 

The 3D covariant derivative has the usual EW structure but with the specific value $\sin\theta_W = 1/2$~\cite{Weinberg:1971nd,Mulders:2016qve}. For bosons, the covariant derivative leads to massive $W^\pm$ and $Z^0$ bosons and a massless photon ($\gamma$) in the usual way, with at zeroth order additional relations like $M = M_{\rm top} = M_{\rm Higgs}\sqrt{2} = 2\,M_Z$. 

The color-space duality (1D to 3D transition) also implies the emergence or freezing of physical gluons in QCD. For hadrons as 3D entangled quark composites, emergence of gluons also appears at high energies ('small x' physics) where staple-like Wilson loops emerge from lightlike Wilson lines and are associated with zero momentum gluons~\cite{Boer:2016xqr}. 

\section{Conclusions}

In this letter, concepts of QIT have been used to construct a minimal Hilbert space incorporating the space-time and SM symmetries.
It can provide a natural framework and a new look at the SM, its many parameters and its open issues such as confinement. The latter is actually not an issue because quarks and leptons just live in different entanglement classes, electroweakly interacting via the bipartite entanglement level. Hadrons being quark composites but also composites such as positronium, atoms and nuclei involve additional entanglement which for hadrons has been indicated in Fig.~\ref{tripartite-physics}. Since the underlying Hilbert space of the tripartite space corresponds to a d = 1+1 field theory, one also has all the advantages of better convergence and absence of naturalness problems~\cite{Stojkovic:2014lha,Calcagni:2009kc,Calcagni:2010bj} 

The next step will be to test this scheme and its implications. Within the SM it will not invalidate the QCD part, nor the EW part by itself, but the hope is that by realizing an underlying simpler basis, one could constrain masses and couplings in the SM and handle situations in which both electroweak and strong aspects come together. At this point no attempts have been made to go beyond the zeroth order, which is certainly necessary. After all, the weak mixing angle deviates from $\sin\theta_W = 1/2$, the mass relations are not precise, the scale of QCD is not clear and the mass is clearly not concentrated in one family. The mass spectrum of quarks and leptons does require a precise mixing of the quantum states. On the other hand, there appear to be possibilities to implement basic SUSY, family structure, linking of space and color into classes of entangled states including the emergence of all symmetries of the SM. Underlying the hadronic structure there is a quantum world that at several levels may give insight into effective descriptions like soft collinear effective theories (SCET, see e.g.\ Ref~\cite{Becher:2014oda}), color glass condensates (see e.g.\ Ref~\cite{Gelis:2010nm}, the AdS/QCD correspondence (see e.g.\ Refs~\cite{deTeramond:2008ht,Brodsky:2016yod}), color-kinematic dualities~\cite{Bern:2008qj} or chiral dynamics at low energies. Even if at this stage, the approach is mostly restricted to a representation of quantum states in the SM, the rich interplay among them make it worthwhile to further pursue these ideas.

\section*{Acknowledgements}

I acknowledge useful discussions with several colleagues at Nikhef and at some workshops among them `POETIC 2015' in Paris, `The proton mass 2017' at ECT* in Trento, `Lightcone 2017' in Mumbai and `New Frontiers in QCD 2018' at the Yukawa Institute for Theoretical Physics in Kyoto. I also want to acknowledge work done on some group theoretical aspects by my MSc student Fabian Springer. 
This research is part of the FP7 EU "Ideas" programme QWORK (Contract 320389).

\bibliographystyle{apsrev}

\end{document}